\documentclass[aps, showpacs, showkeys,nofootinbib,floatfix]{revtex4}

\usepackage{graphicx}
\usepackage{hyperref}

\begin{document}

\title{Reconstruction of Five-dimensional Bounce cosmological Models From Deceleration Factor}

\author{Lixin Xu}
\email{lxxu@dl.cn}
\author{Hongya Liu\footnote{Corresponding author}}
\email{hyliu@dlut.edu.cn}
\author{Yongli Ping}

\affiliation{Department of Physics, Dalian University of
Technology, Dalian, 116024, P. R. China}

\begin{abstract}
In this paper, we consider a class of five-dimensional Ricci-flat
vacuum solutions, which contain two arbitrary functions $\mu(t)$
and $\nu(t)$. It is shown that $\mu(t)$ can be rewritten as a new
arbitrary function $f(z)$ in terms of redshift $z$ and the $f(z)$
can be determined by choosing particular deceleration parameters
$q(z)$ which gives early deceleration and late time acceleration.
In this way, the $5D$ cosmological model can be reconstructed and
the evolution of the universe can be determined.
\end{abstract}


 \hfill TP-DUT/2006-1

\maketitle

\section{Introduction}

Recent observations of high redshift Type Ia supernovae reveal
that our universe is undergoing an accelerated expansion rather
than decelerated expansion \cite{RS,TKB,Riess}. In addition, the
discovery of Cosmic Microwave Background (CMB) anisotropy on
degree scales together with the galaxy redshift surveys indicate
$\Omega _{total}\simeq 1$ \cite{BHS} and $\Omega _{m}\simeq \left.
1\right/ 3$. All these results strongly suggest that the universe
is permeated smoothly by 'dark energy', which violates the strong
energy condition with negative pressure. The dark energy and
accelerating universe has been discussed extensively from
different points of view \cite{Quintessence,Phantom,K-essence}. In
principle, a natural candidate to dark energy could be a small
cosmological constant. However, there exist serious theoretical
problems: fine tuning problem and coincidence problem. To overcome
the fine tuning problem, some self-interact scaler fields $\phi$
with an equation of state (EOS) $w_{\phi}=p_{\phi}\left/
\rho_{\phi}\right.$ were introduced dubbed quintessence, where
$w_{\phi}$ is time varying and negative. In nature, the potentials
of the scalar field would be determined from the underlying
physical theory, such as Supergravity, Superstring/M-theory etc..
However, disregarding these underlying physical theories just from
the phenomenal level, we can design many kinds of potentials to
solve the concrete problems \cite{sahni}. Once the potentials are
given, EOS $w_{\phi}$ of dark energy can be found. On the
contrary, the potential can be reconstructed by a given EOS
$w_{\phi}$ \cite{GOZ}. Then, the forms of the scalar potential are
obtained by the observations data with given EOS $w_{\phi}$.
However, it is mentioned that the same models have early
deceleration epoch. And, the acceleration of the universe just
begins at not distance past. So, starting from the deceleration
parameter, we can also reconstruct the cosmological models
\cite{Banerjee}. This is the main motivation of this paper.

The idea that our world may have more than four dimensions is due
to Kaluza \cite{Kaluza}, who unified Einstein's theory of General
Relativity with Maxwell's theory of Electromagnetism in a $5D$
manifold. In 1926, Klein reconsidered the idea and treated the
extra dimension as a compact small circle topologically
\cite{Klein}. However, up to now, there is no experimental
evidence and indication for the existence of extra dimensions.
Nevertheless, there is strong theoretical motivation for
considering our world spacetime with more than three spatial
dimension. $M$-theory is a higher dimensional theory with seven
extra spatial dimensions, which try to incorporate quantum gravity
in a consistent way. Recently, in higher dimension frame, brane
world model has been studied extensively. In brane world model,
our four dimensional world is a hypersurface (brane) embedded in a
higher dimensional manifold (bulk) and all forces are confined on
the brane but gravity can propagate in the bulk. Also, it has
interesting consequence in cosmology, for recent review seen
\cite{brane-review}. Also, in higher dimensional frame, the
Space-Time-Matter (STM) theory as a modern Kaluza-Klein theory is
designed to incorporate the geometry and matter by Wesson and his
collaborators \cite{Wesson}. In STM theory, our world is
hypersurface embedded in five-dimensional Ricci flat ($R_{AB}=0$)
manifold and all the matter in our world are induced from the
higher dimension. The theory requires the metric components
containing the extra dimension, saying $y$. And, the
compactification or not of the extra dimension is not necessary in
STM theory, for Ricci flat ($R_{AB}=0$) five-dimensional manifold.
The consequent cosmology in STM theory is studied in \cite{LiuW},
\cite{STM-cosmology}, \cite{WLX}.

\section{Dark Energy in a class of five-dimensional cosmological model}

Within the framework of STM theory, a class of exact $5D$
cosmological solution was given by Liu and Mashhoon in 1995
\cite{Liu}. Then, in 2001, Liu and Wesson \cite{LiuW} restudied
the solution and showed that it describes a cosmological model
with a big bounce as opposed to a big bang. The $5D$ metric of
this solution reads
\begin{equation}
dS^{2}=B^{2}dt^{2}-A^{2}\left(
\frac{dr^{2}}{1-kr^{2}}+r^{2}d\Omega ^{2}\right) -dy^{2}
\label{5-metric}
\end{equation}
where $d\Omega ^{2}\equiv \left( d\theta ^{2}+\sin ^{2}\theta
d\phi ^{2}\right) $ and
\begin{eqnarray}
A^{2} &=&\left( \mu ^{2}+k\right) y^{2}+2\nu y+\frac{\nu
^{2}+K}{\mu ^{2}+k},
\nonumber \\
B &=&\frac{1}{\mu }\frac{\partial A}{\partial t}\equiv
\frac{\dot{A}}{\mu }. \label{A-B}
\end{eqnarray}
Here $\mu =\mu (t)$ and $\nu =\nu (t)$ are two arbitrary functions
of $t$, $k$ is the $3D$ curvature index $\left(k=\pm 1,0\right)$,
and $K$ is a constant. This solution satisfies the 5D vacuum
equation $R_{AB}=0$. So, the three invariants are
\begin{eqnarray}
I_{1} &\equiv &R=0, I_{2}\equiv R^{AB}R_{AB}=0,
\nonumber  \\
I_{3} &=&R_{ABCD}R^{ABCD}=\frac{72K^{2}}{A^{8}}. \label{3-invar}
\end{eqnarray}
The invariant $I_{3}$ in Eq. (\ref{3-invar}) shows that $K$
determines the curvature of the 5D manifold. It would be pointed
out that the $5D$ and $4D$ Planck mass are not related directly,
because of $^{5}G_{AB}=0$, in the STM theory. So, in $4D$ we can
take $\kappa_4^2=8\pi G_4$, where $G_4$ is $4D$ Newtonian
gravitation constant.

Using the $4D$ part of the $5D$ metric (\ref{5-metric}) to
calculate the $4D$ Einstein tensor, one obtains
\begin{eqnarray}
^{(4)}G_{0}^{0} &=&\frac{3\left( \mu ^{2}+k\right) }{A^{2}},
\nonumber \\
^{(4)}G_{1}^{1} &=&^{(4)}G_{2}^{2}=^{(4)}G_{3}^{3}=\frac{2\mu \dot{\mu}}{A%
\dot{A}}+\frac{\mu ^{2}+k}{A^{2}}.  \label{einstein}
\end{eqnarray}
In our previous works \cite{WLX}, the induced matter was set to be
a conventional matter plus a time variable cosmological `constant'
or three components: dark matter radiation and $x$-matter. In
\cite{XuL}, we pointed out the correspondence between the
arbitrary function $f(z)$ and scalar field potentials. In this
paper, we assume that the induced matter contains two parts, for
simplicity: cold dark matter (CDM) $\rho_{cd}$ and dark energy
(DE) $\rho_{de}$. So, we have
\begin{eqnarray}
\frac{3\left( \mu ^{2}+k\right) }{A^{2}} &=&\rho_{cd}+\rho_{de},  \nonumber \\
\frac{2\mu \dot{\mu}}{A\dot{A}}+\frac{\mu ^{2}+k}{A^{2}}
&=&-\left(p_{cd}+p_{de}\right), \label{FRW-Eq}
\end{eqnarray}
where
\begin{equation}
p_{cd}=0, \quad p_{de}=w_{de}\rho_{de}. \label{EOS-X}
\end{equation}
From Eqs.(\ref{FRW-Eq}) and (\ref{EOS-X}), one obtains the EOS of
the dark energy
\begin{equation}
w_{de}=\frac{p_{de}}{\rho_{de}}=-\frac{2\left. \mu
\dot{\mu}\right/ A \dot{A}+\left. \left( \mu ^{2}+k\right) \right/
A^{2}}{3\left. \left( \mu ^{2}+k\right) \right/
A^{2}-\rho_{cd0}A^{-3}},\label{wx}
\end{equation}
and the dimensionless density parameters
\begin{eqnarray}
\Omega _{cd} &=&\frac{\rho _{cd}}{\rho_{cd}+\rho_{de}}=\frac{\rho
_{cd0}}{3\left( \mu ^{2}+k\right) A},  \label{omega-cd} \\
\Omega_{de} &=&1-\Omega_{cd}. \label{omega-de}
\end{eqnarray}
where $\rho _{cd0}=\bar{\rho}_{cd0}A_{0}^{3}$, and $\Omega_{cd}$
and $\Omega_{de}$ are dimensionless density parameters of CDM and
DE respectively. The Hubble parameter and deceleration parameter
should be given as \cite{LiuW}, \cite{WLX},
\begin{eqnarray}
H&\equiv&\frac{\dot{A}}{A B}=\frac{\mu}{A} \\
q \left(t, y\right)&\equiv&\left.
-A\frac{d^{2}A}{d\tau^{2}}\right/\left(\frac{dA}{d\tau}\right)^{2}
=-\frac{A \dot{\mu}}{\mu \dot{A}}, \label{df}
\end{eqnarray}
from which we see that $\dot{\mu}\left/\mu\right.>0$ represents an
accelerating universe, $\dot{\mu}\left/\mu\right.<0$ represents a
decelerating universe. So the function $\mu(t)$ plays a crucial
role of defining the properties of the universe at late time.
%%%%%%%%%%%%%%%%%%%%%%%%%%%%%%%%%%%%%%%%%%%%%%%%%%%%%%%%%%%%%%%
In this paper, we consider the spatially flat $k=0$ cosmological
model. From the above Eqs. (\ref{wx})-(\ref{df}), it is easy to
find that the equations does not contain $\nu(t)$ explicitly,
which is included in $A$. So, to avoid boring choice of the
concrete forms $\nu(t)$, we use $A_{0}\left/A \right.=1+z$ and
define $\mu_{0}^{2}\left/ \mu_{z}^{2}\right.=f\left(z\right)$, and
then find that the Eqs. (\ref{wx})-(\ref{df}) can be reduced to as
follows in redshift $z$
\begin{eqnarray}
w_{x} &=&-\frac{1+\left(1+z\right)d\ln
f\left(z\right)\left/dz\right.}{3-3\Omega_{cd}}, \label{wx-2} \\
\Omega_{cd}&=&\Omega_{cd0}\left(1+z\right)f\left(z\right),\label{omega-cd-2} \\
\Omega_{de}&=&1-\Omega_{cd},\label{omega-de-2}\\
q&=&\frac{1+3\Omega_{x}w_{x}}{2}=-\frac{\left(1+z\right)}{2}\frac{d\ln
f\left(z\right)}{dz}. \label{q}
\end{eqnarray}
From Eq. (\ref{q}), we find that it is an ordinary differential
equation of function $f(z)$ w.r.t. redshift $z$, once one form of
$q(z)$ is given. In Ref. \cite{Banerjee}, the scalar potentials
are constructed from a given deceleration parameter
$q=-1-pa^p/(1+a^p)$, where $p$ is a constant. In our case, with
this spirit, we also can reconstruct the forms of function $f(z)$
from a given concrete form of $q(z)$. As a example, we consider
the following deceleration parameter
\begin{equation}
q(z)=\frac{1}{2}-\frac{\alpha}{(1+z)^{\beta}},\label{qz1}
\end{equation}
where, $\alpha$ and $\beta$ are positive constants determined by
observation values. By this assumption, one obtains $f(z)$
\begin{equation}
f(z)=\frac{C\exp\left(-\frac{2\alpha(1+z)^{-\beta}}{\beta}\right)}{1+z}\label{fz},
\end{equation}
where, $C=\exp\left(2\alpha/\beta\right)$ is an integral constant
determined by $f(0)\equiv 1$. Once $f(z)$ is specified, the
evolution parameters, such as $\Omega_{cd}$ $\Omega_{de}$ and
$w_x$ are obtained at one time. From Eq. (\ref{qz1}), one fine
that
\begin{equation}
q_0=\frac{1}{2}-\alpha.
\end{equation}
The current observation value $q_0=-0.67\pm 0.25$ determines the
parameter $\alpha$. And, the parameter $\beta$ has relation with
the ratio of $\Omega_{cd}$ and $\Omega_{de}$, which is constrained
by early cosmological observations. By set $\alpha=1.17$ and
$\beta=1.7$, the evolution of these parameters is plotted in Fig.
\begin{figure}
\centering
\includegraphics[width=4.5in]{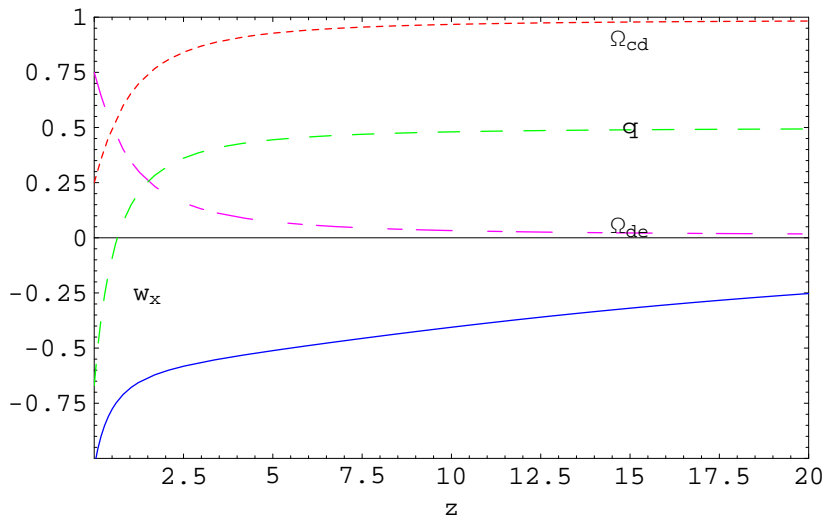}
\caption{The evolution of the dimensionless density parameters
$\Omega_{cd}$, $\Omega_{de}$, and deceleration parameter $q$, EOS
of dark energy $w_x$ versus redshift $z$, where
$\Omega_{cd0}=0.25$, $\Omega_{de 0}=0.75$, $\alpha=1.17$ and
$\beta=1.7$.} \label{fig1}
\end{figure}

\section{conclusions}

A general class of $5D$ cosmological models is characterized by a
big bounce as opposed to the big bang in $4D$ standard
cosmological model. This exact solution contains two arbitrary
functions $\mu(t)$ and $\nu(t)$. By careful observation, one finds
that the dimensionless density, deceleration parameters and EOS of
dark energy are independent on $\nu(t)$ explicitly which is
included in $A$. In terms of redshif $z$, $\mu(t)$ can be
rewritten into a new arbitrary function $f(z)$. Once the forms of
the arbitrary function $f(z)$ are specified, the universe
evolution will be determined. In this paper, we reconstruct the
arbitrary function $f(z)$ by assuming a particular form of
deceleration parameter $q(z)$. In this way, the $5D$ cosmological
models are reconstructed and the evolution of the universe can be
determined.

\acknowledgements{This work was supported by NSF (10273004) and
NBRP (2003CB716300) of P. R. China.}


\begin{thebibliography}{*}
\bibitem{RS} A.G. Riess, et.al., {\it Observational evidence from supernovae for an
accelerating universe and a cosmological constant}, 1998 {\it
Astron. J.} {\bf 116} 1009, astro-ph/9805201; S. Perlmutter,
et.al., {\it Measurements of omega and lambda from 42
high-redshift supernovae}, 1999 {\it Astrophys. J.} {\bf 517} 565,
astro-ph/9812133.

\bibitem{TKB} J.L. Tonry, et.al., {\it Cosmological Results from High-z Supernovae
}, 2003 {\it Astrophys. J.} {\bf 594} 1, astro-ph/0305008; R.A.
Knop, et.al., {\it New Constraints on $\Omega_M$,
$\Omega_\Lambda$, and w from an Independent Set of Eleven
High-Redshift Supernovae Observed with HST}, astro-ph/0309368;
B.J. Barris, et.al., {\it 23 High Redshift Supernovae from the IfA
Deep Survey: Doubling the SN Sample at z>0.7}, 2004 {\it
Astrophys.J.} {\bf 602} 571, astro-ph/0310843.

\bibitem{Riess} A.G. Riess, et.al., {\it Type Ia Supernova Discoveries
at $z>1$ From the Hubble Space Telescope: Evidence for Past
Deceleration and Constraints on Dark Energy Evolution},
astro-ph/0402512.

\bibitem{BHS} P. de Bernardis, et.al., {\it A Flat Universe from High-Resolution
Maps of the Cosmic Microwave Background Radiation}, 2000 {\it
Nature} {\bf 404} 955, astro-ph/0004404; S. Hanany, et.al., {\it
MAXIMA-1: A Measurement of the Cosmic Microwave Background
Anisotropy on angular scales of 10 arcminutes to 5 degrees},2000
{\it Astrophys. J.} {\bf 545} L5, astro-ph/0005123; D.N. Spergel
et.al., {\it First Year Wilkinson Microwave Anisotropy Probe
(WMAP) Observations: Determination of Cosmological
Parameters},2003 {\it Astrophys. J.} Supp. {\bf 148} 175,
astro-ph/0302209.

\bibitem{Quintessence} I. Zlatev, L. Wang, and P.J. Steinhardt ,
\textit{Quintessence, Cosmic Coincidence, and the Cosmological
Constant}, 1999 {\it Phys. Rev. Lett.} {\bf 82} 896,
astro-ph/9807002; P.J. Steinhardt, L. Wang , I. Zlatev, {\it
Cosmological Tracking Solutions}, 1999 {\it Phys. Rev.} D {\bf 59}
123504, astro-ph/9812313; M.S. Turner , {\it Making Sense Of The
New Cosmology}, 2002 {\it Int. J. Mod. Phys.} A {\bf 17S1} 180,
astro-ph/0202008; V. Sahni , {\it The Cosmological Constant
Problem and Quintessence}, 2002, {\it Class.Quant.Grav.} {\bf 19}
3435, astro-ph/0202076.

\bibitem{Phantom} R.R. Caldwell, M. Kamionkowski,
N.N. Weinberg, {\it Phantom Energy: Dark Energy with w $<-1$
Causes a Cosmic Doomsday}, 2003 {\it Phys. Rev. Lett.} {\bf 91}
071301, astro-ph/0302506; R.R. Caldwell , {\it A Phantom Menace?
Cosmological consequences of a dark energy component with
super-negative equation of state}, 2002 {\it Phys. Lett.} B {\bf
545} 23, astro-ph/9908168; P. Singh, M. Sami, N. Dadhich, {\it
Cosmological dynamics of a phantom field}, 2003 {\it Phys. Rev.} D
{\bf 68} 023522, hep-th/0305110; J.G. Hao, X.Z. Li , {\it
Attractor Solution of Phantom Field}, 2003 {\it Phys.Rev.} D {\bf
67} 107303, gr-qc/0302100.

\bibitem{sahni} V. Sahni, {\it Theoretical models of dark
energy}, 2003 {\it Chaos. Soli. Frac.} {\bf 16} 527.

\bibitem{K-essence} Armend\'{a}riz-Pic\'{o}n, T. Damour, V. Mukhanov,
{\it k-Inflation}, 1999 {\it Physics Letters} B {\bf 458} 209; M.
Malquarti, E.J. Copeland , A.R. Liddle, M. Trodden, {\it A new
view of k-essence}, 2003 {\it Phys. Rev.} D {\bf 67} 123503; T.
Chiba , {\it Tracking k-essence}, 2002 {\it Phys. Rev.} D {\bf 66}
063514, astro-ph/0206298.

%%%%%%%%%%%%%%%%%%%%%%%%%%%%%%%%%%%%%%%%%%%%%%%%%%%%%%
\bibitem{GOZ} Z.K. Guo, N. Ohtab and Y.Z. Zhang, {\it Parametrization of Quintessence
and Its Potential}, astro-ph/0505253.

\bibitem{Banerjee} N. Banerjee, S. Das, {\it Acceleration of the universe with a simple
trigonometric potential}, astro-ph/0505121.

\bibitem{Kaluza} T. Kaluza, {\it On The Problem Of Unity In Physics}, Sitzungsber. Preuss. Akad. Wiss.
Berlin (Math. Phys.) K1 966(1921).

\bibitem{Klein} O. Klein, {\it Quantum Theory And Five-Dimensional Relativity}, Z. Phys. 37 895(1926)
[Surveys High Energ. Phys. 5 241(1926)].

\bibitem{brane-review} R. Maartens, {\it Brane-world gravity}, Living Rev.Rel. 7 7(2004),
gr-qc/0312059.

\bibitem{Wesson} P.S. Wesson {\it Space-Time-Matter} (Singapore:
World Scientific) 1999; J.M. Overduin and P.S. Wesson,
\textit{Phys. Rept.} \textbf{283}, 303 (1997), gr-qc/9805018.

\bibitem{LiuW} H.Y. Liu and P.S. Wesson, {\it Universe models with a variable
cosmological ``constant'' and a ``big bounce''}, 2001 {\it
Astrophys. J.} {\bf 562} 1, gr-qc/0107093;

\bibitem{STM-cosmology} T. Liko, P.S. Wesson,
gr-qc/0310067; S.S. Seahra, P.S. Wesson, {\it Universes encircling
five-dimensional black holes}, 2003 {\it J. Math. Phys.}, {\bf 44}
5664; Ponce de Leon J, 1988 {\it Gen. Relativ. Gravit.} {\bf 20}
539; L.X. Xu , H.Y. Liu, B.L. Wang, {\it Big Bounce singularity of
a simple five-dimensional cosmological model}, 2003 {\it Chin.
Phys. Lett.} {\bf 20} 995, gr-qc/0304049; H.Y. Liu, {\it Exact
global solutions of brane universe and big bounce}, 2003 {\it
Phys. Lett.} B {\bf 560} 149, hep-th/0206198;

\bibitem{WLX} B.L. Wang, H.Y. Liu, L.X. Xu, {\it Accelerating Universe in a Big Bounce
Model},Mod. Phys. Lett. A {\bf 19} 449(2004), gr-qc/0304093; L.X.
Xu, H.Y. Liu, {\it Three Components Evolution in a Simple Big
Bounce Cosmological Model}, to appear in IJMPD, astro-ph/0412241.

\bibitem{XuL} L.X. Xu and H.Y. Liu, {\it The Correspondence Between a Five-dimensional Bounce cosmological
Model and Quintessence Dark Energy Models}, have been submitted.

\bibitem{Liu} H.Y. Liu and B. Mashhoon, {\it A machian interpretation of the
cosmological constant}, 1995 {\it Ann. Phys}. {\bf 4} 565.

\bibitem{SHEM} S. Hannestad and E. Mortsell, Phys. Rev. {\bf D66} (2002) 063508.

\bibitem{ARCDH} A.R. Cooray and D. Huterer, Astrophys. J. {\bf 513} (1999) L95.

\bibitem{EVL} E.V. Linder, Phys. Rev. Lett. {\bf 90} (2003) 091301.

\bibitem{TP} T. Padmanabhan and T.R. Choudhury, Mon. Not. Roy. Astron. Soc. {\bf 344} (2003) 823.

\bibitem{BFGGE} B.F. Gerke and G. Efstathiou, Mon. Not. Roy. Astron. Soc. {\bf 335} (2002) 33.

\bibitem{Ponce} J. Ponce de Leon, Gen. Relativ. Gravit. 20 (1988) 539.

\bibitem{Compbell} S.S. Seahra and P.S. Wesson, {\it Application of the Campbell-Magaard
theorem to higher-dimensional physics, }2003 {\it Class. Quant.
Grav.} {\bf 20} 1321, gr-qc/0302015.

\bibitem{Anderson} E. Anderson, {\it The Campbell--Magaard Theorem is inadequate
and inappropriate as a protective theorem for relativistic field
equations}, gr-qc/0409122.

\bibitem{Dahia} F. Dahia and C. Romero, {\it Dynamically generated embeddings of
spacetime}, gr-qc0503103.

\end{thebibliography}
\end{document}